\documentclass[referee,a4paper,12pt,traditabstract]{swsc} 

\usepackage{amsmath}
\usepackage{graphicx}
\usepackage{txfonts}
\usepackage{subfigure}
\usepackage{epstopdf}
\usepackage[authoryear,round]{natbib}
\usepackage[backref]{hyperref}
\usepackage{url}

\hypersetup{colorlinks=true,citecolor=cyan,urlcolor=cyan,linkcolor=blue}

\newcommand{\lambdaoneau}{{\lambda_{\parallel}^{*}}}
\newcommand{\sigmaphimax}{\sigma_{\phi,\, \mathrm{max}}}

\begin{document}


   \title{The effect of turbulence strength on meandering field lines and Solar Energetic Particle event extents}
%
%

   \subtitle{}
   
   \titlerunning{Turbulence strength and SEP cross-field transport}

   \authorrunning{Laitinen et al.}

\author{T. Laitinen
  \inst{1}
  \and
  F. Effenberger
  \inst{2,3}
  \and
  A. Kopp
  \inst{4}
  \and
  S. Dalla
  \inst{1}
}

\institute{Jeremiah Horrocks Institute, University of Central Lancashire,
  Preston, UK\\
  \email{\href{mailto:tlmlaitinen@uclan.ac.uk}{tlmlaitinen@uclan.ac.uk}}
  \and
International Space Science Institute, Bern, Switzerland
\and
Bay Area Environmental Research Institute, Petaluma, CA, USA
  \and
  Universit\'e Libre de Bruxelles, Service de Physique Statistique et des Plasmas,
  CP 231, 1050 Brussels, Belgium
}


 
\abstract{Insights into the processes of Solar Energetic Particle
  (SEP) propagation are essential for understanding how solar
  eruptions affect the radiation environment of near-Earth space. SEP
  propagation is influenced by turbulent magnetic fields in the solar
  wind, resulting in stochastic transport of the particles from their
  acceleration site to Earth. While the conventional approach for SEP
  modelling focuses mainly on the transport of particles along the
  mean Parker spiral magnetic field, multi-spacecraft observations
  suggest that the cross-field propagation shapes the SEP fluxes at
  Earth strongly. However, adding cross-field transport of SEPs as
  spatial diffusion has been shown to be insufficient in modelling the
  SEP events without use of unrealistically large cross-field
  diffusion coefficients. Recently, Laitinen et al.\ (2013b, 2016)
  demonstrated that the early-time propagation of energetic particles
  across the mean field direction in turbulent fields is not
  diffusive, with the particles propagating along meandering field
  lines. This early-time transport mode results in fast access of the
  particles across the mean field direction, in agreement with the SEP
  observations. In this work, we study the propagation of SEPs within
  the new transport paradigm, and demonstrate the significance of
  turbulence strength on the evolution of the SEP radiation
  environment near Earth. We calculate the transport parameters
  consistently using a turbulence transport model, parametrised by the
  SEP parallel scattering mean free path at 1~AU, $\lambdaoneau$, and
  show that the parallel and cross-field transport are connected, with
  conditions resulting in slow parallel transport corresponding to
  wider events. We find a scaling $\sigmaphimax\propto
  (1/\lambdaoneau)^{1/4}$ for the Gaussian fitting of the longitudinal
  distribution of maximum intensities. The longitudes with highest
  intensities are shifted towards the west for strong scattering
  conditions. Our results emphasise the importance of understanding
  both the SEP transport and the interplanetary turbulence conditions
  for modelling and predicting the SEP radiation environment at
  Earth.}


\maketitle

\section{Introduction}

Solar Energetic Particles (SEPs), accelerated in solar eruptive
events, pose a significant space weather threat to man-made technology
and astronauts \citep{spaceradrisk}. To forecast SEP fluxes near
Earth's orbit, it is important to understand how their acceleration is
related to flares, coronal mass ejections and other related phenomena
during solar eruptions. Furthermore, as the particles propagate
through a turbulent solar wind medium, predicting the fluxes and
fluences at 1~AU requires understanding of how the solar wind
turbulence affects the charged particle motion.
%
%

The propagation of SEPs in a turbulent medium is typically modelled as
random walk due to the stochastic nature of magnetic field
fluctuations, and described as spatial and velocity diffusion using a
Fokker-Planck formalism \citep{Parker1965,Jokipii1966}. The
propagation along the mean-field is usually modelled as either spatial
or pitch angle diffusion \citep{Jokipii1966}. The cross-field
transport, on the other hand, is usually described as spatial
diffusion due to random walk of the turbulent magnetic field lines
\citep{Jokipii1966}, compounded by the parallel scattering
\citep{Matthaeus2003,Shalchi2010a,RuffoloEa2012}. These approaches
have support in full-orbit particle simulations \citep{GiaJok1999} and
galactic cosmic ray observations
\citep{Burger2000,Potgieter2014}. However, several recent
observational studies suggest faster propagation of SEPs
across the mean field than predicted by the current theoretical
understanding: they often require a ratio of the cross-field
diffusion coefficient to the parallel one of order
$\kappa_\perp/\kappa_\parallel\sim 0.1-1$
\citep{Zhang2003,Dresing2012,Droge2014} \footnote{note, however,
  \citet{Droge-2016} who obtained $\kappa_\perp/\kappa_\parallel\sim
  0.02$ in some cases} , whereas values $\kappa_\perp/\kappa_\parallel\lesssim 0.01$
are more consistent with the interplanetary turbulence conditions at
1~AU \citep{Burlaga1976,Pei2011difftens,LaEa2016parkermeand}.

Recently, \citet[L2013 in the following]{LaEa2013b} demonstrated, using full-orbit particle
simulations in turbulent magnetic fields superposed on a constant
background magnetic field, that SEPs can propagate rapidly to large
cross-field distances along turbulently meandering field-lines already
early in SEP event history. While the concept of field-line meandering
is included in earlier models in the diffusion coefficient,
L2013 showed that the initial SEP cross-field transport is
non-diffusive, and cannot be modelled using a diffusion approach.  As
further shown in \citet{LaDa2017decouple}, the particles remain on
their initial meandering field lines up to tens of hours before
decoupling and spreading more freely across the meandering field
lines. Thus, the initial evolution of SEP events is dominated by
systematic widening of the SEP cross-field distribution, while
diffusion dominates the evolution of the SEP cross-field distribution
only tens of hours after the SEP injection. L2013 pointed
out also that the early-time non-diffusive SEP propagation across the
mean field direction is much faster than the time-asymptotic
cross-field diffusion.

Using the novel modelling approach introduced in L2013,
\citet{LaEa2016parkermeand} developed a particle transport model in
the heliospheric Parker Spiral magnetic field configuration. They
demonstrated that in moderate turbulence conditions, parametrised by
the parallel scattering mean free path $\lambda_\parallel=$~0.3~AU,
fast spreading of SEPs across the field to a wide range of longitudes,
as found by multi-spacecraft observations of SEP events
\citep[e.g.][]{Lario2006,Dresing2012,Richardson2014,Dresing2014},
could be obtained already with a narrow source region.

\citet{LaEa2016parkermeand} considered a case study of SEP propagation
in magnetic turbulence characterised by the value of the parallel
scattering mean free path at 1~AU, $\lambdaoneau\equiv
\lambda_\parallel(r=1\mathrm{ AU})=$~0.3~AU, whereas in reality the
turbulence, and as a consequence the particle transport parameters,
can vary considerably from event to event
\citep{Burlaga1976,Bavassano1982JGR,Palmer1982,WaWi93}, and even be
different during an event at different heliographic longitudes
\citep{Droge-2016}. Using full-orbit simulations with a constant
background magnetic field, \citet{LaDa2017decouple} and
\citet{LaEa2017meandstatistic} showed that the initial cross-field
extent of the SEP distribution depends strongly on the turbulence
amplitude. Thus, it is important to evaluate the effect of different
levels of turbulence amplitudes on the SEP event width in the Parker
Spiral geometry. In this study, we compare longitudinal SEP event
extents for different levels of plasma
turbulence, as parametrised by parallel mean free paths
$\lambdaoneau$. We concentrate on 10~MeV protons, which have received less attention in multi-spacecraft-observed SEP event modelling. While electrons and protons are often considered to be accelerated in different processes and source regions \citep[e.g.][]{Reames1999}, the recently observed similar heliolongitudinal extents for electrons and protons during different events \citep[e.g.][]{Richardson2014} may suggest similar processes responsible for the spreading of these particles in interplanetary space, warranting closer analysis of their cross-field transport.
We present the employed models in Section~\ref{sec:Models}, the
results in Section~\ref{sec:results} and discuss and draw our
conclusions in Sections~\ref{sec:discussion}
and~\ref{sec:conclusions}.

\section{Models}\label{sec:Models}

The Fokker-Planck \& Field Line Random Walk (FP+FLRW) model used in this study is based on the findings of L2013, who used full-orbit particle simulations in turbulent magnetic fields to show that the initial cross-field propagation of charged particles in turbulent magnetic fields is non-diffusive. The particles tend to follow their field lines, which spread across the mean field direction due to turbulent fluctuations. Until a particle decouples from its field line, its propagation across the mean background magnetic field is deterministic, in the sense that particles which scatter in their pitch angle from one pitch angle hemisphere to the other will just retract their path along the same stochastically meandering path. 
Thus, the particle cross-field transport behaviour remains non-Markovian at times shorter than the timescale over which the particle decouples from its original field line. L2013, and subsequently \citet{LaDa2017decouple} and \citet{LaEa2017meandstatistic} showed that this non-Markovian propagation can dominate an SEP event for up to tens of hours, depending on the turbulence conditions. The slow onset of the decoupling of particles from the meandering field lines can explain the intensity dropouts observed in some SEP events \citep[e.g.][]{Mazur2000}, as shown for example in simulations by \citet{Tooprakai2016}.

While the motion resulting from field-line meandering and field-parallel scattering has been described as compound (sub)diffusion by earlier researchers \citep[e.g.][]{Kota2000}, L2013 discovered that during early times, the particle propagation cannot be described as diffusion at all, as the particles retain memory of their propagation history.

L2013 introduced the FP+FLRW model, where this process is described as combination of particle propagation along meandering path (supplemented with pitch angle scattering) and particle diffusion across the meandering field. As shown in L2013 and subsequently further investigated in \citet{LaEa2017meandstatistic}, the model agrees well with full-orbit simulations at early times, when the particles are still tied to their field lines, and at the time-asymptotic limit, where the particle cross-field propagation is fully diffusive.

In the full-orbit simulations in L2013, the meandering of field lines was created by using a superposition of Fourier modes corresponding to spectra of slab- and 2D-mode waves, which can be cumbersome particularly in scenarios more complex than the constant background field used in that work. For simpler and faster Fokker-Planck description of particle propagation, the FP+FLRW mode considers a description of diffusively meandering particle paths, based on results of, e.g., \citet{Matthaeus1995}. In the FP+FLRW approach, rather than calculating the complete fluctuating magnetic field, the effect of diffusive spreading field lines on the propagating particles is estimated by propagating each particle on a separately drawn stochastic meandering path. Thus, schematically the algorithm of the FP+FLRW model for each pseudo-particle in the simulations proceeds as follows:
\begin{enumerate}
\item[1] Calculate a diffusively meandering path, unique to the particle being simulated.
\item[2] Propagate a pseudo-particle until end of simulation time using the following scheme:
  \begin{enumerate}
  \item Step along the meandering path.
  \item Diffuse and adiabatically focus the pitch angle. Here, for the focusing the mean background magnetic field is used.
  \item Take a diffusive cross-field spatial step in direction perpendicular to the meandering path.
  \end{enumerate}
\end{enumerate}

It should be emphasised that in the FP+FLRW model each simulated pseudo-particle has only one meandering field line, and the particle diffuses across this meandering field line: The pseudo-particle does not switch from one meandering path to another. The individual particles propagating each at their unique meandering paths facilitate the initial non-diffusive evolution of the particle distribution seen in L2013 and \cite{LaEa2017meandstatistic}, whereas the particle's spatial cross-field diffusion from this meandering path facilitates the time-asymptotic diffusive particle transport.

An alternative approach, with a particle changing from one meandering path to another, could also be devised. However, such a model would require precise description of the decoupling of particles from their initial fieldlines, and the relation of that decoupling process to the turbulent field line separation, which are not yet well understood. Thus, as the simpler FP+FLRW model was shown by L2013 to reproduce the full-orbit particle simulations well, the single-meandering-path approach is well-justified.

The stochastically meandering path is described as field-line diffusion using a stochastic differential equation \citep[SDE,][]{Gardiner2009,Strauss-Effenberger-2017}, with the displacement $dr_\perp$ across the Parker field  direction given as

\begin{equation} \label{eq:fldiff}
  dr_\perp=\sqrt{2 D_{FL}(r)\,dr_\parallel}W_\perp,
\end{equation}
where $dr_\parallel$ is a step along the local Parker spiral
direction, and $W_\perp$ a Gaussian random number with zero mean and
unit variance. The field-line diffusion coefficient $D_{FL}(r)$ is
calculated based on \citet{Matthaeus1995}, using the
  radially-evolving 2D component of the turbulence spectrum discussed
below.

This method of calculating the meandering field line using a stochastic method is naturally statistic in nature, and does not reproduce patchy spatial particle distributions seen in some full-orbit particle simulations \citep[e.g.][]{Tooprakai2016}, which may explain intensity dropouts observed in SEPs \citep[e.g.][]{Mazur2000} \citep[see also discussion in ][]{LaDa2017decouple}. It also cannot replicate the coherence of nearby field lines, but, as shown in \citep{Ruffolo2004}, such coherence is lost in 2D-dominated turbulence at scales which are small compared to heliospheric scales.

The particle propagation along meandering field lines, the step B in the FP+FLRW scheme, is performed using an SDE formulation of the Fokker-Planck equation \citep{Roelof1969,Skilling1971,Isenberg1997,Zhang2009,Strauss-Effenberger-2017} . The Fokker-Planck equation is given as

\begin{align}
    \frac{\partial f}{\partial t}
+ \left(\mu v\mathbf{b}_m+\mathbf{V}_{sw}\right)\cdot\nabla f
+ \frac{v}{2L}(1-\mu^2)\frac{\partial f}{\partial \mu}\nonumber\\
+\left[\frac{\mu(1-\mu^2)}{2}\left(\nabla\cdot\mathbf{V}_{sw}-3\mathbf{b}\mathbf{b}:\nabla\mathbf{V}_{sw}\right)\right] \frac{\partial f}{\partial \mu}
 \nonumber\\ 
+\left[\frac{1-3\mu^2}{2}
\mathbf{b}\mathbf{b}:\nabla\mathbf{V}_{sw}
-\frac{1-\mu^2}{2}\nabla\cdot\mathbf{V}_{sw}\right]p\frac{\partial f}{\partial
p}\nonumber\\
= \frac{\partial}{\partial \mu} \left( D_{\mu\mu}\frac{\partial
    f}{\partial \mu}\right)+\nabla\cdot\hat\kappa\nabla f  +Q(\mathbf{r},\mathbf{v},t),\label{eq:FPE}
\end{align}
where $v$ and $\mu$ are the particle's velocity and pitch-angle cosine, respectively, and $Q$ is the particle source function.  $\mathbf{V}_{sw}$  is the solar wind velocity, $\mathbf{b}$ a unit vector along the Parker magnetic field $\mathbf{B}$, and $\mathbf{b}_m$ a unit vector along the meandering path given by Eq.~(\ref{eq:fldiff}). The focusing length $L=-B/( \partial B/\partial s)$, with $s$ the arc-length along the field-line, is calculated using the mean Parker spiral field. The particles scatter in pitch angle cosine $\mu=v_\parallel/v$ according to the pitch-angle diffusion coefficient $D_{\mu\mu}$, and across the mean field direction according to the spatial cross-field diffusion coefficient $\kappa_{xx}=\kappa_{yy}\equiv\kappa_\perp$, the non-zero elements of the cartesian diffusion tensor $\hat\kappa$. In this study, we ignore the energy changes given by the 5th term in Equation~(\ref{eq:FPE}), as they are expected to be small during the early phase of the SEP event \citep[e.g.][]{Dalla2015}. The remaining equation is solved using the SDE code described in further detail in \citet{Kopp2012}.

The particles are propagated along a path that consists of a
Parker spiral field superposed with stochastic fluctuations, resulting in particle paths that meander about the Parker spiral. The magnitude of the magnetic field is taken as the mean Parker spiral field value,

\begin{equation}
  B(r)=B_0\left(\frac{r_0}{r}\right)^2 \sqrt{\frac{r^2+a^2}{r_0^2+a^2}},
\end{equation}
where $B_0=5$~nT is the magnetic field strength at
$r_0=1$~AU, and $a=V_{sw}/(\Omega_\odot\sin\theta)$, where
$V_{sw}=400$~km/s, $\Omega_\odot=2.8631\cdot10^{-6}$~rad/s is the
solar rotation rate and $\theta$ the co-latitude.

As shown by L2013 and \citet{LaDa2017decouple}, at intermediate timescales, of the order of the parallel scattering timescale of the particles, the particles begin to decouple from their field lines and eventually
time-asymptotically approach diffusive cross-field propagation. We
include the transition to the time-asymptotic cross-field diffusion into our simulations by diffusing the particles across
the meandering field. While this approach is not precise, L2013
demonstrated that it is much more accurate than using
only cross-field diffusion from the mean field, or only particle
propagation along meandering field lines. The cross-field diffusion
coefficient $\kappa_\perp$ is calculated using the Non-Linear Guiding Centre theory
\citep[NLGC,][]{Matthaeus2003} for the spectrum described below. We do not incorporate the recently suggested dependence of $\kappa_\perp$ on the particle's pitch angle \citep[][]{Droge2010,QinShalchi2014,Strauss2015} since full-orbit results \citep{LaDa2017decouple} indicate that it might be more complicated than the suggested proportionality to $|\mu|$ or $1-\mu^2$ .

It should be noted that although both the meandering path of the particle and the cross-field diffusion coefficient are calculated from the same turbulence spectrum, this does not amount to taking the effect of meandering field lines on particles into account twice. As discussed in \citet{LaDa2017decouple}, the initial cross-field spreading of the particles is caused by the particles following the meandering field lines. The cross-field diffusion, on the other hand is caused by particles decoupling from the field lines and following new field lines, which meander relative to the original field line \citep[see also][]{RuffoloEa2012}. Thus, the two phenomena, while both related to field-line meandering, are separate and must be both accounted for.

Finally, the particles also scatter as they propagate along the
meandering field line. We model this by using a quasi-linear
pitch-angle diffusion coefficient $D_{\mu\mu}$ \citep[e.g.][]{Jokipii1966}, with additional scattering at $\mu=0$ to close the resonance gap, as suggested by \citet{BeeckWibberenz1986} (see \citet{LaEa2016parkermeand} for details).

The particle and field line diffusion coefficients are calculated
using a heliospheric turbulence spectrum with slab and 2D components
\citep{Gray1996}. The turbulence spectrum is given as

\begin{equation}
  \label{eq:spectrum}
  S(\mathbf{k})\equiv S(\mathbf{k},r_0)=S_\perp(k_\perp,r_0)\delta(k_\parallel)+S_\parallel(k_\parallel,r_0)\delta(\mathbf{k}_\perp),
\end{equation}
where $k_\parallel$ and $k_\perp=\left|\mathbf{k}_\perp\right|$ are
the parallel and perpendicular wavenumbers, and
$S_\parallel(k_\parallel)$ and $S_\perp(k_\perp)$ are broken power law
spectra as given in \cite{LaEa2016parkermeand}. It should be noted that our turbulence model differs from the one introduced by \citet{Giacalone2001}, in which the turbulence is generated by motion of magnetic field footpoints due to solar supergranulation. The latter does not allow for further turbulence evolution of the magnetic fields in interplanetary space \citep[e.g.][and references therein]{Bruno2013}, and thus limits the meandering of interplanetary field lines to the angular scale of the supergranular motion.

We model the radial evolution of the turbulence within the
heliosphere using the WKB transport approximation
\citep{Richter1974,TuPuWei1984}. For simplicity, we neglect wave
refraction, changes in the wave geometry and the modulus of $k$, as
well as non-linear evolution of the spectral shape \citep[see][for
discussion]{LaEa2016parkermeand}. We further consider constant radial
solar wind velocity $V_{sw,r0}$, and electron density $
n_e(r)=n_{e0}\;r_0^2/r^2$. With these assumptions, the radial
evolution of the turbulence spectrum can be written as

\begin{equation}\label{eq:wkb}
 S_{\parallel,\perp}(k_{\parallel,\perp},r)= S_{\parallel,\perp}(k_{\parallel,\perp},r_0) \left(\frac{r_0}{r}\right)^3\left(\frac{V_{sw,r0}+v_{A0}}{V_{sw,r0}+\frac{r_0}{r}v_{A0}}\right)^2,
\end{equation}
where $V_{sw,r0}=400$~km/s is the constant solar wind velocity, and
subscript 0 denotes the values at reference distance $r_0$, and
$v_{a,r0}=30$~km/s is the Alfv\'en velocity at $r_0=1$~AU. The resulting $\propto r^3$ trend of the wave power is consistent with turbulence observations \citep[e.g.][]{Bavassano1982JGR}.

The spectral power of the slab and 2D components, $   S_{\parallel,\perp}(k_{\parallel,\perp},r_0)$, is parametrised by   the total turbulence amplitude $\delta B^2=2\int   S(\mathbf{k})d\mathbf{k}$, and the energy ratio $\delta   B_\parallel^2/\delta B_\perp^2$ between the slab and 2D modes, for   which we use $20\%:80\%$, as suggested by \citet{Bieber1996}. The total   turbulence amplitude is parametrised by the parallel mean free path at 1~AU, $\lambdaoneau\equiv \lambda_\parallel(r=1 \mathrm{AU})$, as given by the quasilinear theory \citep{Jokipii1966} for the slab  spectrum $S_{\parallel}(k_{\parallel})$ at 1~AU. It should be   emphasised that the parallel mean free path is fixed using   $\lambdaoneau$ only at 1~AU: elsewhere all particle and field line   transport parameters are calculated consistently using the   turbulence model given by Eqs.~(\ref{eq:spectrum})   and~(\ref{eq:wkb}). Thus, we constrain the radial evolution of both   parallel and perpendicular transport parameters consistently,   instead of using an ad hoc parametrisation.

\begin{figure}
   \centering
   \includegraphics[width=.95\textwidth]{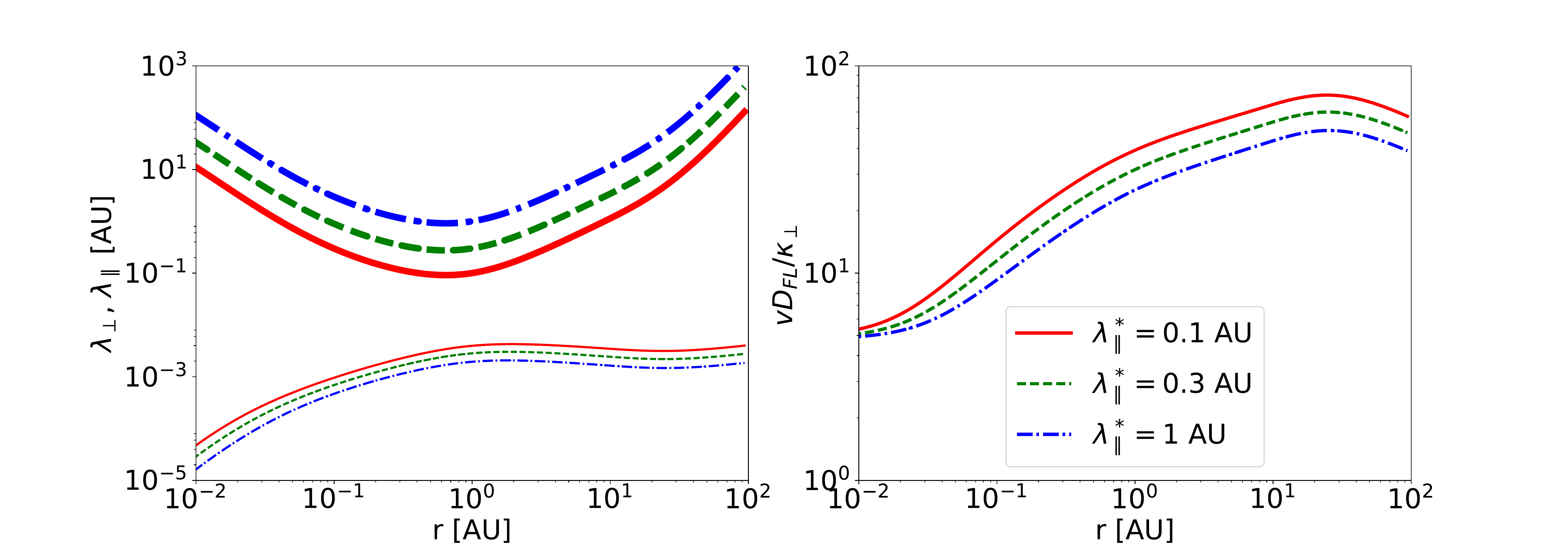}
   \caption{Left panel: Parallel (thick curves) and perpendicular
     (thin curves) particle mean free paths as function of radial
     distance from the Sun, for 10 MeV protons and different
     turbulence strengths parametrised by $\lambdaoneau$. Right
     panel: The ratio of the field line and cross-field particle
     diffusion coefficients.}\label{fig:kappas}
\end{figure}

The parallel and cross-field mean free paths for 10~MeV protons are shown as a function of distance from the Sun in the left panel of Figure~\ref{fig:kappas} for the modelled turbulence corresponding to $\lambdaoneau$ values of 0.1, 0.3 and 1~AU. Close to the Sun, the parallel mean free path is large, indicating nearly scatter-free propagation, and decreases to the parametrised value $\lambdaoneau$ at 1~AU, after which it increases again. On the other hand, the cross-field mean free path is very short close to the Sun and increases initially faster than $\propto r$, indicating that the diffusion coefficient ratio $\kappa_\perp/\kappa_\parallel$ is not constant in the heliosphere, but depends strongly on the radial distance from the Sun. Similar results of the radial dependence of the particle diffusion coefficients have recently been presented in several studies \citep{Pei2011difftens,Strauss2017perpel,Chhiber2017CRdiffGlobmodel}.

In the right panel of Figure~\ref{fig:kappas}, we describe how the
cross-field particle and field line diffusion coefficients evolve in
the heliosphere, by presenting their ratio as a function of radial
distance from the Sun. As discussed in L2013, particles
propagate initially along meandering field lines that spread
diffusively according to diffusion coefficient
$D_{FL}$. Time-asymptotically, the cross-field propagation is
diffusive, described by the particle cross-field diffusion coefficient
$\kappa_\perp$, which is much slower than the spreading of particles
non-diffusively along the field lines, due to particles scattering
along the meandering field lines. As can be seen in the right panel of
Figure~\ref{fig:kappas}, the spreading of particles across the field
due to the early process, at rate $v D_{FL}$, is 1-2 orders of
magnitude faster than the time-asymptotic diffusive cross-field
spreading of the particles, and the difference increases as a function
of distance from the Sun. The ratio $v D_{FL}/\kappa_\perp$ decreases
for weaker turbulence (larger $\lambdaoneau$), and, as discussed in
\citet{LaEa2016parkermeand}, $v D_{FL}$ and
$\kappa_{\perp}$ calculated using the NLGC \citep{Matthaeus2003}
converge to the same value in the limit of no parallel scattering for
a particle beam.

\section{Results}\label{sec:results}

\begin{figure}
   \centering
   \includegraphics[width=\textwidth]{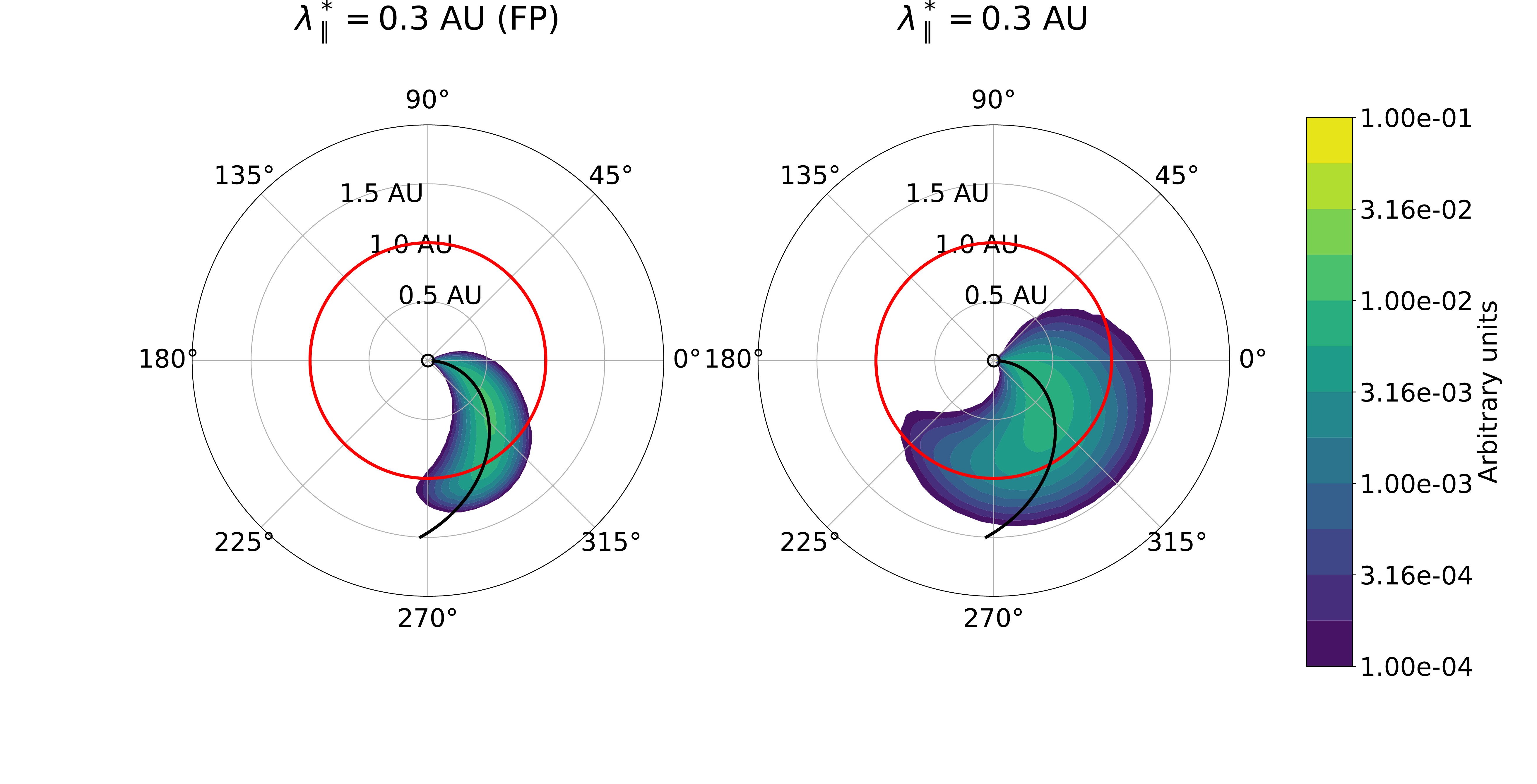}
   \caption{Distribution of 10 MeV SEPs integrated between latitudes $-10^\circ$ 
     and $10^\circ$, in arbitrary units, 2
     hours after impulsive injection at
     $(r,\phi,\theta)=(1\,\mathrm{r}_\odot,0,\pi/2)$, for the FP model
     (left panel) and the FP+FLRW model (right panel), respectively,
     with $\lambdaoneau=0.3 \mathrm{AU}$. The red curve depicts 1~AU
     radial distance, and the thick black spiral curve the Parker
     field connected to the injection location.}\label{fig:fpvsfpflrw}
\end{figure}

\begin{figure}
   \centering
   \includegraphics[width=\textwidth]{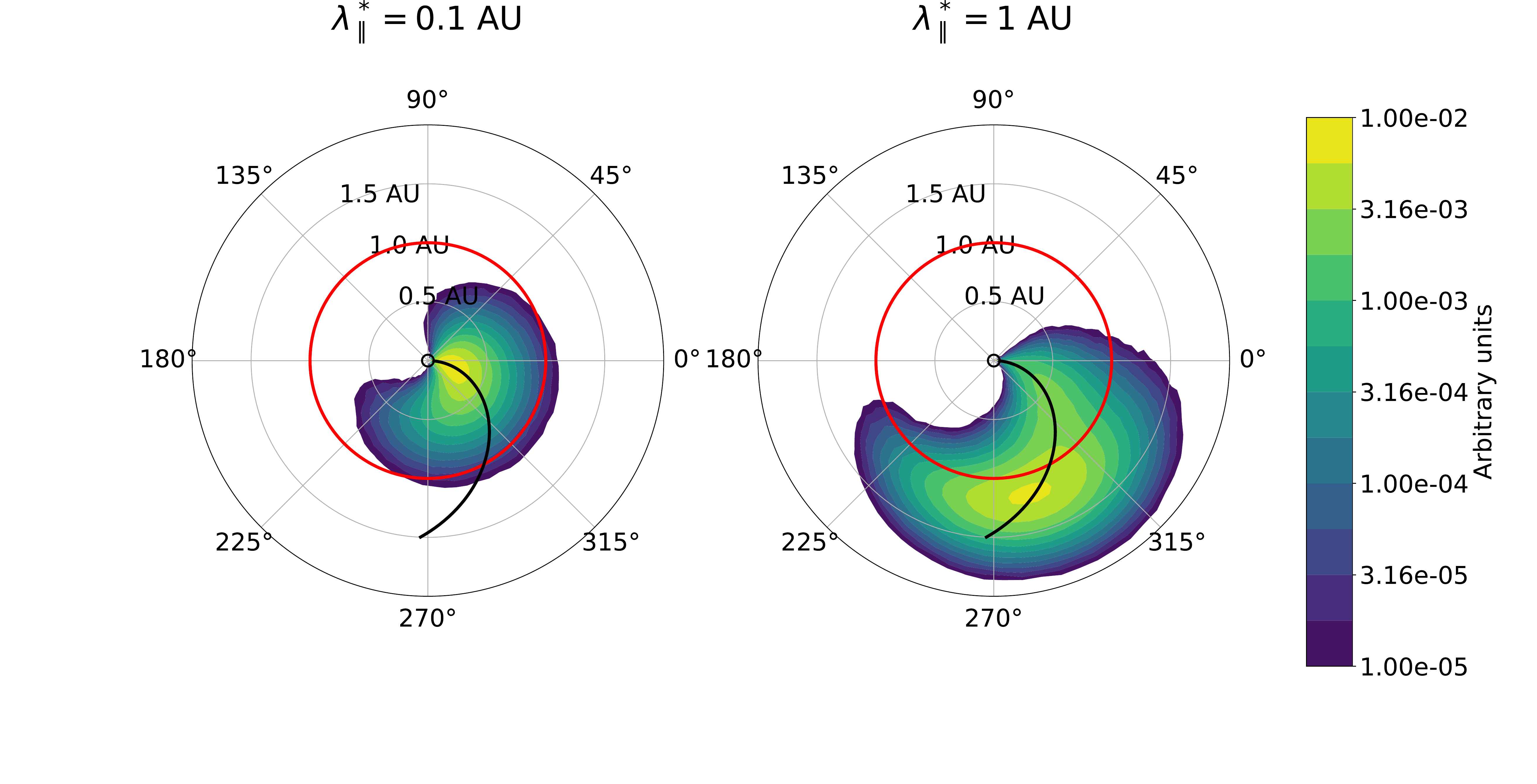}
   \caption{Distribution of 10 MeV SEPs integrated between latitudes $-10^\circ$ 
     and $10^\circ$, as in Fig.~\protect\ref{fig:fpvsfpflrw},
     with left panel for FP+FLRW with $\lambdaoneau=$0.1~AU, and
     right panel for FP+FLRW with
     $\lambdaoneau=1$~AU.}\label{fig:parkerlambdacomp}
\end{figure}

We study the effect of turbulence strength on SEP event evolution in time, both along and across the mean field direction. We use a simple injection profile

\begin{equation}
  \label{eq:inj}
  Q(r,\theta,\phi,t)=\delta(r-1 \mathrm{r}_\odot) \delta(\theta-\pi/2)
  \delta(\mu-1)\delta(E-E_0)\delta(\phi) \delta(t),
\end{equation}
where $(r,\theta,\phi)$ define the spherical coordinate system
$\mathrm{r}_\odot$ is the solar radius, and $E_0=10$~MeV the energy of
the simulated protons. The coronal magnetic field can be
  complicated, varying considerably from event to event. However, in
  this study we are interested in SEP propagation in general, instead
  of during a particular SEP event. For this reason, we model the coronal magnetic field  simply as a Parker spiral starting from the
  injection height at $1~\mathrm{r}_\odot$, reserving case studies that investigate the spatial structure of the source region of particles for
  future work.

The results of our study can be extended to other injection profiles
by simply convolving the impulse response with more complicated
injection profiles. It should be noted that \citet{Strauss2017perpel}
demonstrated recently that the source size at or near the Sun
plays only a minor role in the cross-field extent of an SEP event in cases where
cross-field propagation of particles is efficient. Thus, our results
can be considered to represent the SEP event evolution as injected
from a narrow to an intermediate-size SEP source region.

We first show an overview of the early SEP event extent in
Figures~\ref{fig:fpvsfpflrw} and~\ref{fig:parkerlambdacomp}, as a
snapshot of the spatial distribution of 10 MeV protons in the inner
heliosphere, two hours after the injection. The red circle depicts
Earth's orbit, and the black spiral curve the Parker spiral connected
to the injection location. The SEP distribution is given as a function
of heliolongitude and the heliocentric radial distance, integrated
over latitudes $\pm 10^\circ$.

In Figure~\ref{fig:fpvsfpflrw}, we present the SEP distributions for
turbulence parametrised with $\lambdaoneau=0.3$~AU. The left panel
depicts the model where the field-line meandering is omitted
\citep[the FP model in][]{LaEa2016parkermeand}, whereas the right
panel is obtained from the model described in
Section~\ref{sec:Models}. As discussed in \citet{LaEa2016parkermeand},
the primary effect of including the field line meandering into the
modelling is that the particles spread rapidly across the mean Parker
field direction to a wide range of heliolongitudes, as compared to the
slow spreading of the SEPs across the mean field depicted in the left
panel of Figure~\ref{fig:fpvsfpflrw}.

In Figure~\ref{fig:parkerlambdacomp}, we show the effect of changing the turbulence strength on the radial and cross-field extent of an SEP event. In the left panel, the turbulence amplitude has been increased to result in stronger parallel scattering conditions, as parametrised by $\lambdaoneau=$0.1~AU. The differences with the $\lambdaoneau=$0.3~AU case in the right panel of Figure~\ref{fig:fpvsfpflrw} are notable. The strong parallel scattering of the SEPs prevents the particles from propagating as far into the heliosphere as in the $\lambdaoneau=$0.3~AU case. On the other hand, the core of the SEP distribution in the left panel of Figure~\ref{fig:parkerlambdacomp} is considerably wider. This is caused by the particles following the meandering field lines which diffuse with $D_{FL}\propto\delta B/B$ in
2D-dominated turbulence \citep{Matthaeus1995}. Thus, while particle propagation along the mean field line is inhibited by strong scattering in stronger turbulence, the cross-field transport of the SEP distribution is enhanced by the stronger meandering of the field lines.

For weaker turbulence (the $\lambdaoneau=1$~AU case), presented in the right panel of Figure~\ref{fig:parkerlambdacomp}, we see that the reduced parallel scattering of the SEPs causes an increased radial extent of the particle distribution: the front of the SEP population has propagated to nearly 2 AU from the Sun, consistent with nearly-scatter-free propagation of 10 MeV protons of $\sim1$~AU/h. On the other hand, the population is narrower in the cross-field direction, as compared to the cases with $\lambdaoneau=$0.3~AU and 0.1~AU. This is due to the reduced turbulent meandering of the field lines in weaker turbulence.

\begin{figure}
   \centering
   \includegraphics[width=.9\textwidth]{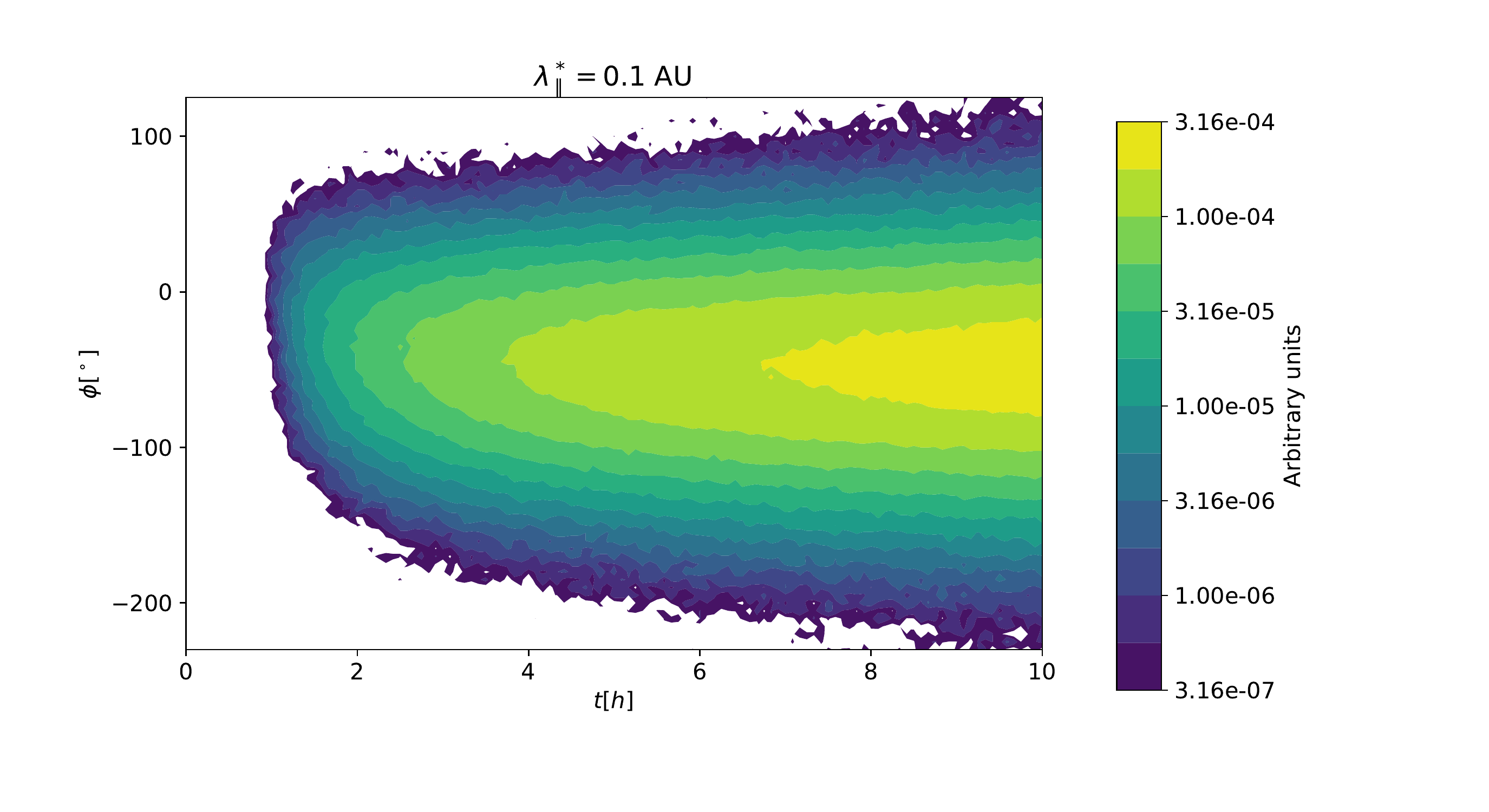}
   \caption{10 MeV SEP intensity at 1 AU, as a function of time and heliographic longitude, with magnetic connection along the Parker spiral at $\phi=-62^\circ$, for $\lambdaoneau=0.1$~AU.\label{fig:phitime_lambda01}}
\end{figure}

\begin{figure}
   \centering
   \includegraphics[width=.9\textwidth]{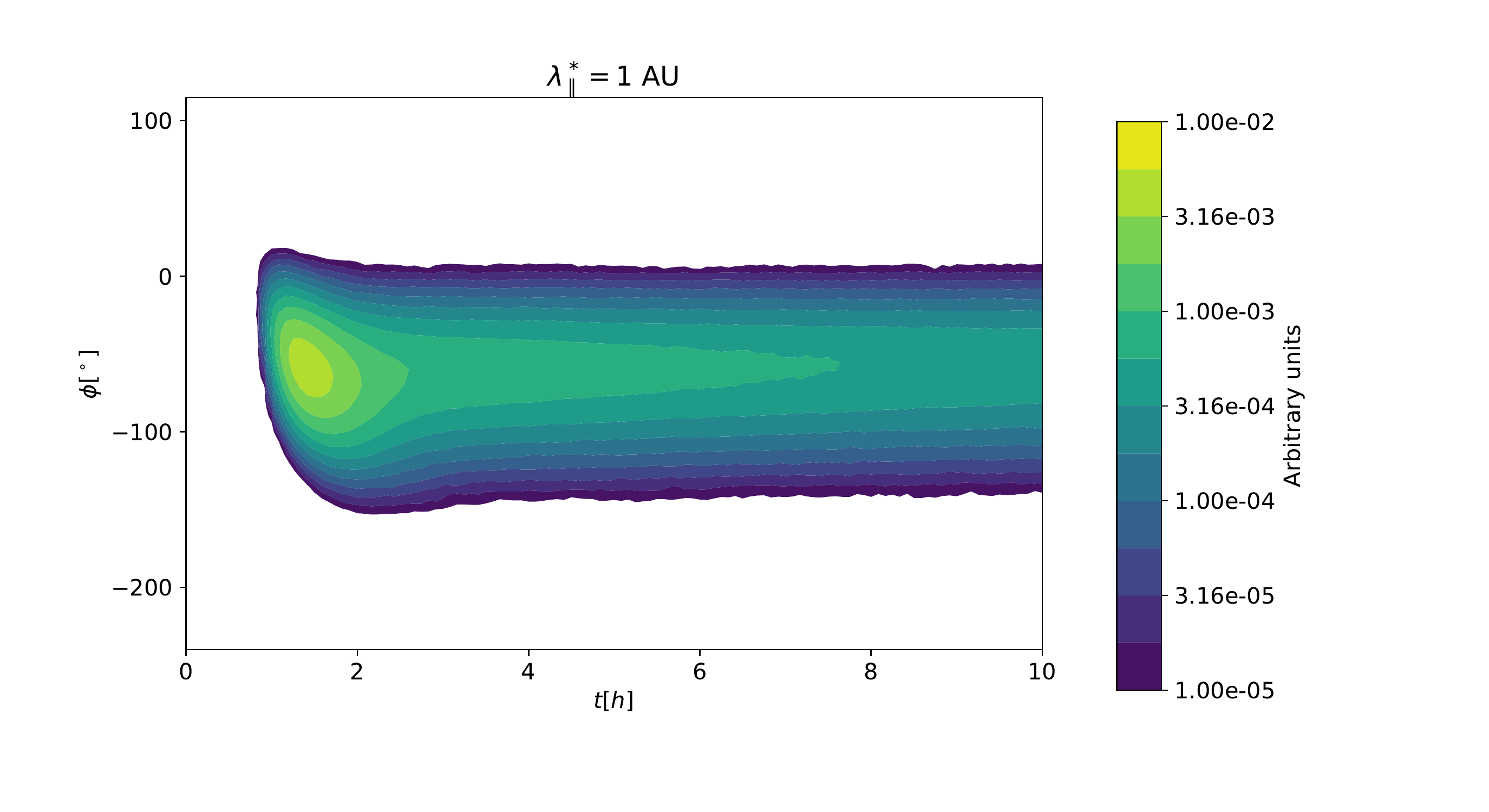}
   \caption{10 MeV SEP intensity at 1 AU, as a function of time and heliographic longitude, with magnetic connection along the Parker spiral at $\phi=-62^\circ$, for $\lambdaoneau=1$~AU.\label{fig:phitime_lambda1}}
\end{figure}

In Figures~\ref{fig:phitime_lambda01} and \ref{fig:phitime_lambda1},
we show the evolution of the SEP event at 1~AU as a function of
heliographic longitude and time, for $\lambdaoneau=$0.1~AU and 1~AU,
respectively. In both cases, the particles spread rapidly across the
mean field direction, with the onset seen at a wide range of
longitudes within the first 2 hours of the event. In the strong
scattering case (Figure~\ref{fig:phitime_lambda01}), the diffusive
nature of the cross-field propagation of the particles after the
initial fast spreading results in the parabole shape of the
high-intensity contours as a function of time and longitude, which
suggest the diffusive scaling of the longitudinal variance of the
particles as $\sigma_{\phi}^2(t)\propto t$ after the initial fast
expansion along the meandering field lines.

The low scattering case (Figure~\ref{fig:phitime_lambda1}) is very
different from the stronger scattering case shown in
Figure~\ref{fig:phitime_lambda01}. The proton intensity increases
rapidly to its maximum value during the first two hours from the
injection, and then begins to decay. This is due to the particles
being nearly scatter-free and focusing adiabatically outwards when
they first arrive to 1~AU. The longitudinal width of the particle
distribution is almost completely determined by the diffusive
spreading of the field lines: there is no appreciable additional
longitudinal widening of the particle distribution after the first two
hours during the simulation period. This is most likely due to a
combination of two effects: The cross-field particle diffusion
coefficient for the case $\lambdaoneau=1$~AU is half of that of the
case $\lambdaoneau=0.1$~AU, thus less cross-field spreading of the
particles can be expected. In addition, the particles escape from the
inner heliosphere very efficiently due to adiabatic focusing and weak
parallel scattering. Thus, the widening of the SEP distribution due to
cross-field propagation of SEPs is compensated by the escape of
particles to the outer heliosphere, resulting in almost constant
intensity at longitudes far from the longitude $\phi=-62^\circ$
connected to the SEP source along the Parker spiral.

\begin{figure}
   \centering
   \includegraphics[width=.9\textwidth]{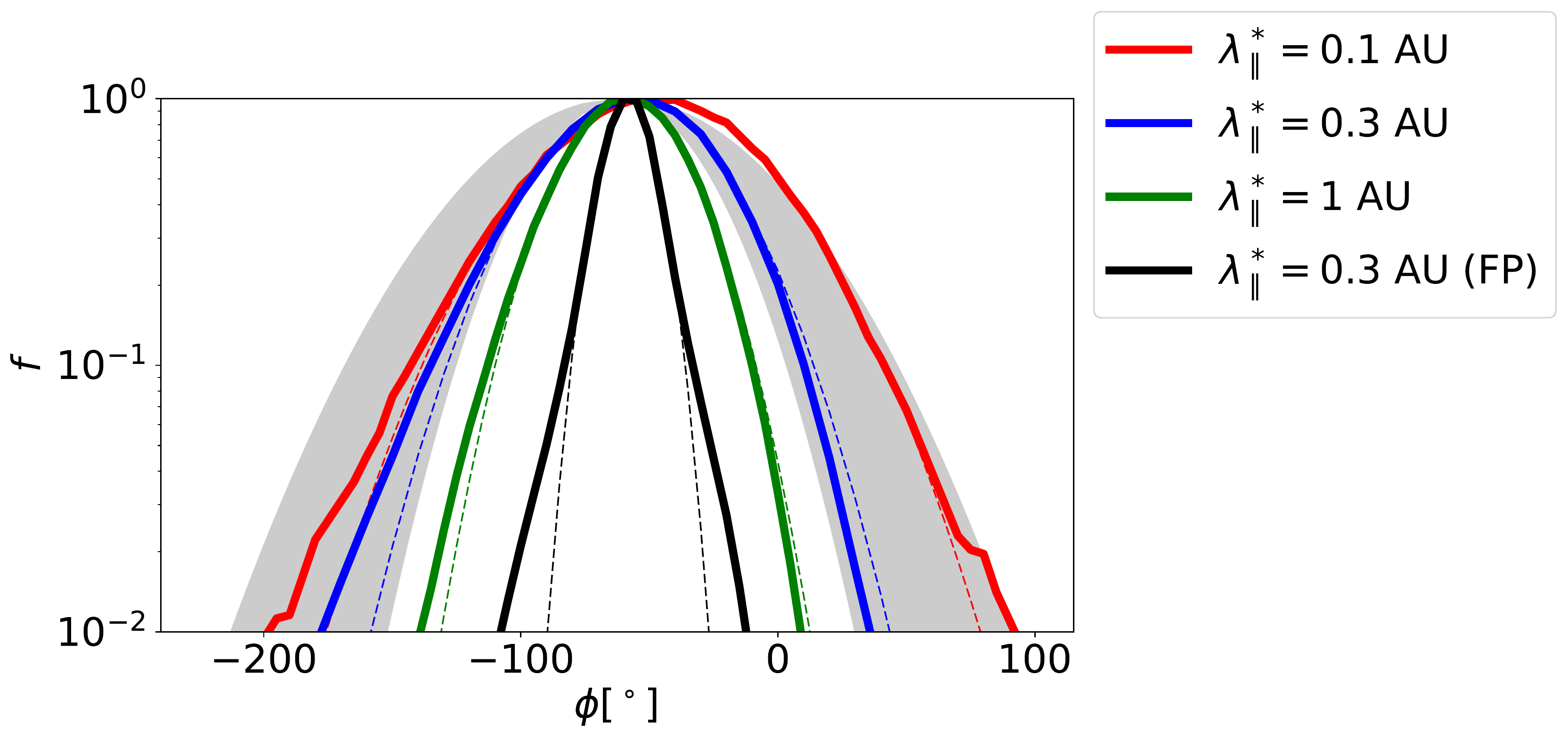}
   \caption{The 10 MeV SEP maximum intensity at 1~AU during the first
     10 hours of the event, as a function of longitude. The thin
     dashed curves show the fitted Gaussian profiles with
     $\sigmaphimax^2=41^\circ$, $33^\circ$, $23^\circ$ for
     $\lambdaoneau=$0.1~AU (red curve), 0.3~AU (blue curve) and 1~AU
     (green curve), respectively, and $10^\circ$ for the reference FP
     simulation case with $\lambdaoneau=$0.3~AU (black curve). The
     gray area depicts the observational range
     $\sigmaphimax=30^\circ-50^\circ$.\label{fig:phimax}}
\end{figure}

The longitudinal extent of the SEP events, as observed using multiple
spacecraft observations, is typically quantified by fitting a Gaussian
curve to the observed peak intensities at different
longitudes. Several case and statistical studies report the standard
deviation $\sigmaphimax$ of the Gaussian to be in the range of
$30^\circ-50^\circ$ for both electrons and ions at different energies, in both
gradual \citep{Dresing2012,Lario2013,Richardson2014,Dresing2014} and impulsive SEP events \citep{Wiedenbeck2013,Cohen2014}. We
present the longitudinal distribution of the peak intensities during
the first 10~hours for our simulation cases in
Figure~\ref{fig:phimax}, including the observational range
$\sigmaphimax=30^\circ-50^\circ$ shown with the gray area, and the
conventional Fokker-Planck result for $\lambdaoneau=$0.3~AU with the black curve.

\begin{figure}
   \centering
   \includegraphics[width=.9\textwidth]{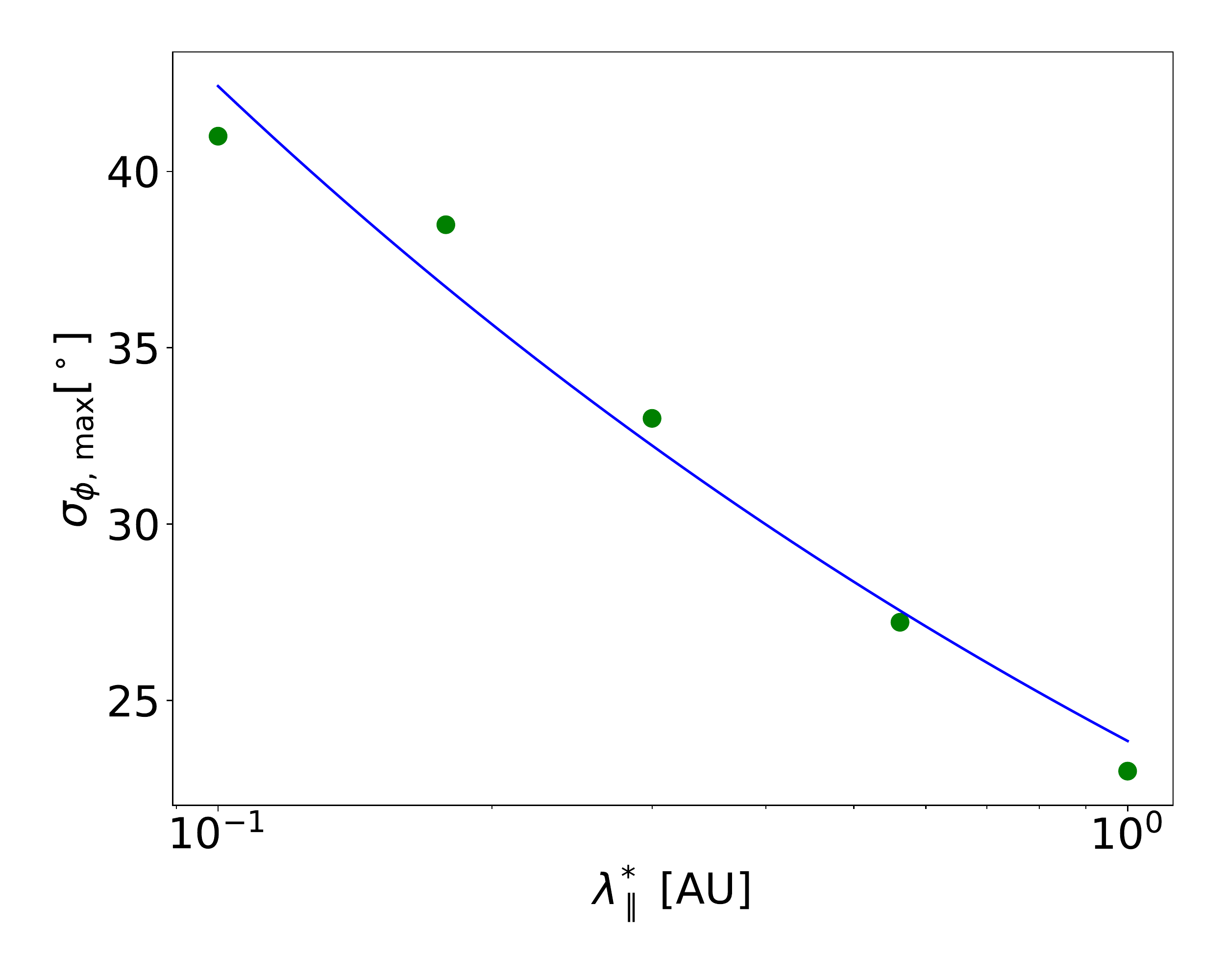}
   \caption{The Gaussian $\sigmaphimax$ fitted to the 10 MeV SEP
     maximum intensity at 1~AU during the first 10 hours of the event,
     as a function of mean free path (green circles). The blue curve
     depicts the trend $\sigmaphimax\propto \left(1/\lambdaoneau\right)^{1/4}$
     expected for particles propagating along meandering field lines.\label{fig:lambdavssigma}}
\end{figure}

As can be seen, the longitudinal width of the SEP event at 1~AU
depends on the turbulence strength, with strong turbulence resulting
in considerably wider SEP events, with $\sigmaphimax=41^\circ$ for the
$\lambdaoneau=$0.1~AU case, as compared to the narrow
$\sigmaphimax=23^\circ$ for the $\lambdaoneau=$1~AU case. We
demonstrate the dependence of $\sigmaphimax$ on the turbulence
amplitude, as parametrised by $\lambdaoneau$, in
Figure~\ref{fig:lambdavssigma}. The blue curve shows the expected
trend for SEPs propagating solely along meandering field lines, which
implies $\sigmaphimax^2\sim D_{FL}\propto\delta B/B$ which, with
$\lambda_\parallel\propto B^2/\delta B^2$ results in $\sigmaphimax\propto \left(1/\lambdaoneau\right)^{1/4}$. As shown in
Figure~\ref{fig:lambdavssigma}, our model results follow this scaling
well. The slight deviation of the expected  $\sigmaphimax\propto \left(1/\lambdaoneau\right)^{1/4}$ trend is likely to be caused by more efficient cross-field diffusion of particles from the meandering field lines by the peak time for small $\lambdaoneau$, as the peak times will be progressively later for smaller $\lambdaoneau$.

Also evident in the longitudinal distribution of the SEPs in
Figure~\ref{fig:phimax} is the asymmetry of the distribution with
respect to the longitude connected to the injection site,
$\phi=-62^\circ$. The centres of the Gaussians are shifted to the
West, with the longitude of the centre of the Gaussian
$\phi_{\mathrm{max}}=-48^\circ$ for the strong turbulence case with
$\lambdaoneau=0.1$~AU. For weaker turbulence cases,
$\phi_{\mathrm{max}}$ approaches the best-connected longitude, with
$-57^\circ$ and $-59^\circ$ for the $\lambdaoneau=0.3$~AU and 1~AU
cases, respectively. Similar shifts were also found in simulations by
\cite{Strauss2015} and \cite{Strauss2017perpel}. A shift of the
maximum of around $10-15^\circ$ to the West has been reported in
multi-spacecraft observed SEP events
\citep{Lario2006,Lario2013,Richardson2014}. It should be noted though
that the multi-spacecraft measurements are typically performed using a
maximum of three measurement points, which makes estimation of the
exact shape and asymmetries of the longitudinal distributions
difficult. In addition, \cite{Richardson2014} reported the centre of
the longitudinal peak distribution as $15^\circ\pm 35^\circ$ west of
the connected longitude, emphasising the large errors associated in
both determining the SEP source location and errors in inferring the
magnetic connection of the observing spacecraft to the source
location.

\section{Discussion}\label{sec:discussion}

Our study shows that meandering field lines are able to efficiently
spread SEPs across the mean Parker Spiral direction at wide range of
heliospheric turbulence conditions, even in weak scattering
conditions. The propagation of SEPs along meandering field lines
results in longitudinally wide SEP events, with dependence of the
longitudinal width scaling with the SEP parallel mean free path as
$\sigmaphimax\propto \left(1/\lambdaoneau\right)^{1/4}$. The evolution
of the SEP event after the initial phase is strongly dependent on the
amount of turbulence in the heliosphere. In strong scattering
environment, the longitudinal extent of the SEPs increases
diffusively, (Figure~\ref{fig:phitime_lambda01}), whereas in the weak
turbulence case, after the initial fast expansion, the longitudinal
extent of the SEP event remains unchanged for the first 10 hours
(Figure~\ref{fig:phitime_lambda1}).

Our results also emphasise the importance of correctly accounting for
the link between the interplanetary turbulence conditions and the
particle transport coefficients. High turbulence amplitudes result in
strong particle scattering along the mean field direction, and hence a
short parallel mean free path \citep[e.g.][]{Jokipii1966}. The
particle propagation across the mean field, on the other hand is more
efficient in stronger turbulence, as shown in both simulation studies
\citep[e.g.][]{GiaJok1999,LaDa2017decouple,LaEa2017meandstatistic} and
theoretical work \citep[e.g.][]{Matthaeus2003,
  Shalchi2010a,RuffoloEa2012}.  This can be clearly seen in
Figure~\ref{fig:kappas}, where the evolution of the parallel and
perpendicular scattering mean free paths are
anticorrelated. Using quasilinear theory and the
  field-line diffusion coefficient from \citet{Matthaeus1995}, $D_{FL}$, we find $D_{FL}\propto
  1/\sqrt{\lambda_\parallel}$, a scaling which is also consistent with
   \citet{Droge-2016} (their Figure~17). Also, changing the geometry of the turbulence from the often-used $\delta B_\parallel^2/\delta B_\perp^2$ energy ratio 20\%:80\% would have an influence on the SEP event evolution: increasing the proportion of the 2D component would result in faster onsets with wider cross-field extents. The overall dependence of $\phi_{\mathrm{max}}(\lambdaoneau)$ shown in Figure~\ref{fig:lambdavssigma} would however, likely stay similar for a fixed $\delta B_\parallel^2/\delta B_\perp^2$ ratio. Other refinements of turbulence modelling, such as incorporating scale-dependence for the $k_\perp/k_\parallel$ anisotropy \citep{GoSr1995,Shalchi2010b,LaEa2013a} and dynamical evolution of the turbulence \citep[accounted for parallel propagation in][]{Bieber1994}, would naturally affect both early and late cross-field evolution of particle populations. These are left for future studies.

  The interdependency between the parallel and cross-field SEP
  transport parameters, is typically ignored in
  parametric 3D SEP transport studies
  \citep[e.g.][]{Zhang2009,Droge2010, He2011, Giajok2012}, and ad-hoc
  values are typically used. While the recent studies by
  \citet{LaEa2016parkermeand} and \citet{Strauss2017perpel} did model
  SEP propagation with consistently modelled SEP transport
  coefficients or one set of turbulence parameters, our paper is to
  our knowledge the first to consistently study the effect of varying
  turbulence strength on both parallel and cross-field propagation
  when modelling SEP events in 3D.

The turbulence parameters are typically observed using in-situ
instruments onboard individual spacecraft, providing a single-point
measurement of the turbulence properties. The SEPs, however, propagate
across the mean field, sampling different heliolatitudes and
longitudes, and their propagation is affected by the 3D turbulent
structure of the heliosphere. Thus, to improve our ability to estimate
the radiation environment in near-Earth space, we should consider a 3D
picture of turbulence in the heliosphere. The need for longitudinally
resolved particle transport conditions was recently highlighted also
by \citet{Droge-2016}, who found that the SEP intensities observed by
the STEREO and ACE spacecraft during a single SEP event may require
different diffusion coefficients for fitting the SEP observations at
different longitudes. Recent work by \citet{Thomas2017} has shown
promise of using solar wind observations at the Lagrangian point L5,
60$^\circ$ behind Earth at Earth's orbit, for forecasting the solar
wind conditions at L1. Such forecast would also make it possible to
evaluate either the average or longitudinally dependent SEP transport
parameters at a wide range of longitudes, from L5 to Earth and
beyond. The recently proposed space weather missions to L5
\citep{Akioka2005L5,Gopalswamy2011_L5,INSTANT2016,CarringtonMission2015}
thus could bring considerable improvement to our ability to model SEP
events and improve our knowledge of the radiation environment in
near-Earth space.

\section{Conclusions}\label{sec:conclusions}

We have studied how the strength of the turbulence in the
interplanetary medium affects SEP event evolution within the new
paradigm introduced by L2013 that includes the early
non-diffusive cross-field transport of SEPs along meandering field
lines. We found that
\begin{itemize}
\item The parallel and cross-field transport of SEPs are inherently
  linked through the turbulence properties, with high levels of
  turbulence resulting in diffusively spreading wide, gradually-rising
  SEP events, and low turbulence in fast SEP events which remain at
  nearly constant longitudinal extent after the initial rapid
  cross-field spreading along meandering field lines.
\item The longitudinal distribution of 10 MeV proton peak intensity
  follows approximately a Gaussian shape, with the longitudinal width
  of the distribution scaling as $\left(1/\lambdaoneau\right)^{1/4}$.
\item In strong turbulence, the longitudinal distribution of the
  particles is asymmetric with respect to the longitude connected to
  the injection site, with the center of the fitted Gaussian
  distribution shifted by $14^\circ$ to the west. Weaker turbulence
  cases are less skewed with respect to the connected longitude.
\end{itemize}

Our results show that knowledge of the turbulence conditions of the
heliospheric plasmas is crucial for modelling the cross-field
propagation of the SEPs early in the events. To forecast the particle
radiation conditions at Earth due to solar eruptions we must
understand the full chain of phenomena including the injection of the
particles at Sun, the physics behind their propagation in the
interplanetary medium, and state of the  interplanetary turbulence
during the SEP propagation within the heliosphere.

\begin{acknowledgements}
  TL and SD acknowledge support from the UK Science and Technology
  Facilities Council (STFC) (grants ST/J001341/1 and ST/M00760X/1),
  and FE from NASA grants NNX14AG03G and NNX17AK25G. The contribution
  of AK benefited from financial support through project He 3279/15-1,
  funded by the Deutsche Forschungsgemeinschaft (DFG), at the CAU
  Kiel, where large parts of this work were carried out. Access to the
  University of Central Lancashire's High Performance Computing
  Facility is gratefully acknowledged. The editor thanks R. Du Toit
  Strauss and an anonymous referee for their assistance in evaluating
  this paper
\end{acknowledgements}


\clearpage
\newpage


\end{document}